\documentclass[preprint,amsmath,amssymb,prd,superscriptaddress]{revtex4}
\usepackage[dvips]{graphicx}
\usepackage{color}
\usepackage{multirow}
\usepackage{array}
\usepackage{amsmath,bm}
\usepackage{booktabs}
\newcommand{\p}{\partial}
\newcommand{\pslash}{p\kern-1ex /}
\newcommand{\lslash}{l\kern-1ex /}
\newcommand{\kslash}{k\kern-1ex /}
\newcommand{\dslash}{\p\kern-1.2ex /}
\newcommand{\Dslash}{{\cal D}\kern-1.5ex /}

\def\Id{\mbox{1\hspace{-1.2mm}I} }

\newcommand{\Dodwf}{\mathcal{D}}
\newcommand{\bea}{\begin{eqnarray}}
\newcommand{\eea}{\end{eqnarray}}

\newcommand{\nn}{\nonumber\\}
\newcommand{\EQ}{\hspace{-2mm} &=& \hspace{-2mm}}

\newcommand{\BAN}{\begin{eqnarray*}}
\newcommand{\EAN}{\end{eqnarray*}}
\def\u{{\bf u}}
\def\d{{\bf d}}

\begin{document}

\newcommand{\CQSE}{
  Center for Quantum Science and Engineering, National Taiwan University, Taipei~10617, Taiwan
}

\newcommand{\NTU}{
  Physics Department, National Taiwan University, Taipei~10617, Taiwan
}

\newcommand{\NTNU}{
  Physics Department, National Taiwan Normal University, Taipei~11617, Taiwan
}

\newcommand{\IoP}{
        Institute of Physics, Academia Sinica, Taipei 11529, Taiwan
}

\preprint{NTUTH-17-505C}

\title{Mass Preconditioning for the Exact One-Flavor Action in Lattice QCD with Domain-Wall Fermion}

\author{Yu-Chih~Chen}
\affiliation{\NTU} 

\author{Ting-Wai~Chiu}
\affiliation{\NTU}
\affiliation{\CQSE}
\affiliation{\NTNU}
\affiliation{\IoP}

\collaboration{TWQCD Collaboration}
\noaffiliation


\begin{abstract}

The mass-preconditioning (MP) technique has become a standard tool to enhance the efficiency   
of the hybrid Monte-Carlo simulation (HMC) of lattice QCD with dynamical quarks, 
for 2-flavors QCD with degenerate quark masses, as well as its extension to the case of one-flavor  
by taking the square-root of the fermion determinant of 2-flavors with degenerate masses.
However, for lattice QCD with domain-wall fermion, the fermion determinant of 
any single fermion flavor can be expressed as a functional integral with  
an exact pseudofermion action $ \phi^\dagger H^{-1} \phi $, 
where $ H^{-1} $ is a positive-definite Hermitian operator without taking square-root, and with 
the chiral structure \cite{Chen:2014hyy}.
Consequently, the mass-preconditioning for the exact one-flavor action (EOFA)  
does not necessarily follow the conventional (old) MP pattern. 
In this paper, we present a new mass-preconditioning for the EOFA, which 
is more efficient than the old MP which 
we have used in Refs. \cite{Chen:2014hyy,Chen:2014bbc}.
We perform numerical tests in lattice QCD with $ N_f = 1 $ and 
$ N_f = 1+1+1+1 $ optimal domain-wall quarks, with one mass-preconditioner 
applied to one of the exact one-flavor actions,   
and we find that the efficiency of the new MP is more than 20\%  
higher than that of the old MP.

\end{abstract}

\maketitle

\section{Introduction}

Consider the action of lattice QCD with one quark flavor 
\BAN
\label{eq:qcd1f}
S = S_g(U) + \bar\psi D(U,m) \psi, 
\EAN
where $ S_g(U) $ is the gauge action in terms of the link variables $ U $, 
$ \bar\psi $ and $ \psi $ are the quark fields, 
and $ D(U,m) $ is the lattice Dirac operator with bare quark mass $ m $, 
satisfying the properties that $ \det D(U,m) > 0 $ for $ m > 0 $, 
and there exists a positive-definite Hermitian Dirac operator $ H(U,m) $ such that $ \det D = \det H $. 
The partition function of this system is 
\bea
\label{eq:Z}
Z = \int dU d\bar\psi \ d\psi \ e^{-S} = \int dU e^{-S_g(U)} \det D(U,m).
\eea
Then the Monte Carlo (MC) simulation of this system amounts to generate a set of configurations
with the probability distribution 
\bea
\label{eq:prob}
e^{-S_g(U)} \det D(U,m),   
\eea
and the quantum expectation value of any physical observable $ T(D^{-1}, U) $
can be obtained by averaging over this set of configurations 
\BAN 
\left< T \right> = \frac{1}{N} \sum_{i=1}^N T(D^{-1}_i, U_i) + {\cal O} \left(\frac{1}{\sqrt{N}}\right),  
\EAN
with the error proportional to $ 1/\sqrt{N} $, where $ N $ is the number of configurations. 

Since the evaluation of $ \det D $ is prohibitively expensive even for small lattices 
(e.g., $ 16^3 \times 32 $), it is common to express $ \det D $ as
\bea
\label{eq:detD}
\det D = \det H = \int d \phi^\dagger d \phi \exp\left( - \phi^\dagger H^{-1} \phi \right),
\eea
where the complex scalar fields $ \phi $ and $ \phi^\dagger $ are called pseudofermion fields, 
carrying the color and Dirac indices but obeying the Bose statistics. 
Then the partition function (\ref{eq:Z}) becomes 
\BAN
\label{eq:Zpf}
Z = \int dU d\phi^\dagger d\phi e^{-S_g[U] - S_{pf}} 
  = \int dU d\phi^\dagger d\phi e^{-S_g[U] - \phi^\dagger H^{-1} \phi }, 
\EAN
where $ S_{pf} = \phi^\dagger H^{-1} \phi $ is called the pseudofermion action. 
Even if the fermion determinant is estimated stochastically with (\ref{eq:detD}), it is still very 
difficult to obtain the desired probability distribution (\ref{eq:prob}) with the conventional 
algorithms (e.g., Metropolis algorithm) in statistical mechanics.
A way out is to introduce a fictituous Hamiltonian dynamics 
with conjugate momentum for each field variable, and to update all fields and momenta globally
followed by a accept/reject decision for the whole configuration, 
i.e., the hybrid Monte-Carlo (HMC) algorithm \cite{Duane:1987de}.
Since the pseudofermion action $ S_{pf} = \phi^\dagger H^{-1} \phi $ is positive-definite, 
$ \phi $ can be generated by the heat-bath method with the Gaussian noise $ \eta $ 
satisfying the Gaussian distribution $ \exp(-\eta^\dagger \eta) $, i.e.,   
to solve the following equation  
\bea
\label{eq:CG_gaussian} 
\frac{1}{\sqrt{H}} \phi = \eta,   
\eea 
where the details of solving (\ref{eq:CG_gaussian}) are suppressed here.
Then the fictituous molecular dynamics only involves the gauge fields 
$ \left\{ A_l \right\} $ and their conjugate momenta $ \left\{ P_l \right\} $,
where $A_l = A_l^a t^a$ is the matrix-valued gauge field corresponding
to the link variable $U_l = \exp(iA_l^a t^a)$.
The Hamiltonian of the molecular dynamics is
\BAN
\mathcal{H}=\frac{1}{2}\sum_{l,a}(P_l^a)^2+S_g[U]+ \phi^\dagger {H}^{-1} \phi, 
\EAN
and the partition function can be written as
\BAN
{\cal Z}=\int[dU][dP][d\phi][d\phi^{\dag}]\exp(-\mathcal{H}).
\EAN
The Hamilton equations for the fictituous molecular dynamics are
\bea
\label{eq:molecular_U}
\frac{dA_l^a(\tau)}{d\tau} \EQ \frac{\partial\mathcal{H}}{\partial P_l^a(\tau)}=P_l^a(\tau)
\Leftrightarrow \frac{dU_l(\tau)}{d\tau}=iP_l(\tau)U_l(\tau), \\
\label{eq:molecular_P}
\frac{dP_l^a(\tau)}{d\tau} \EQ -\frac{\partial\mathcal{H}}{\partial A_l^a(\tau)}
=-\frac{\partial S_g}{\partial A_l^a(\tau)}-\frac{\partial S_{pf}}{\partial A_l^a(\tau)}.  
\eea
These two equations together imply that $ d\mathcal{H}/d\tau = 0 $, which gives
\BAN
\label{eq:molecular_P_a}
P_l^a\frac{dP_l^a(\tau)}{d\tau}=-\frac{dS_g}{d\tau}-\frac{dS_{pf}}{d\tau},
\EAN
as an alternative form of (\ref{eq:molecular_P}). 

The algorithm of HMC simulation can be outlined as follows:
\begin{enumerate}
\item Choose an initial gauge configuration $\{U_l\}$.
\item Generate $P_l^a$ with Gaussian weight $\exp(\{P_l^{a}\}^2/2)$.
\item Generate $\eta$ with Gaussian weight $\exp(-\eta^\dagger\eta)$.
\item Compute $\phi $ according to (\ref{eq:CG_gaussian}).
\item With $\{\phi\}$ held fixed, integrate (\ref{eq:molecular_U}) and (\ref{eq:molecular_P})
      with an algorithm (e.g., Omelyan integrator \cite{Omelyan:2001}) which ensures exact reversibility 
      and area-preserving map in the phase space for any $\delta\tau$.
\item Accept the new configuration $\{U_l' \}$ generated by the molecular dynamics with probability
      $\textmd{min}(1,e^{-\Delta\mathcal{H}})$, 
      where $\Delta\mathcal{H}\equiv\mathcal{H}(U_l^{\prime},P_l^{\prime})-\mathcal{H}(U,P)$.
      This completes one HMC trajectory. 
\item For the next trajectory, go to (2).  
\end{enumerate}

The most computational intensive part of HMC is in the molecular dynamics (MD), the part (5), 
which involves the computation of the fermion force $ -\partial S_{pf}/\partial A_l^a(\tau) $
with the conjugate gradient (CG) algorithm (or other iterative algorithms) 
at each step of the numerical integration in Eq. (\ref{eq:molecular_P}).
Thus, to optimize the efficiency of HMC is to minimize the total computational cost of a MD trajectory 
while making $ \Delta\mathcal{H}$ small enough such that a high acceptance rate can be maintained.
Since $ \Delta\mathcal{H}$ depends on the order of the integrator and 
the size of the integration step $ \epsilon = 1/n $ 
(assuming the total time of the MD trajectory is equal to 1), 
a good balance between the computational cost and the discretization error 
is the Omelyan integrator \cite{Omelyan:2001}.
Moreover, since the fermion force is much smaller than
the gauge force $ -\partial S_g/\partial A_l^a(\tau) $, it is feasible  
to turn on the fermion force less frequent than the gauge force, resulting   
the multiple-time scale (MTS) method \cite{Sexton:1992nu} which speeds up MD significantly. 
Besides MTS, mass preconditioning (MP) \cite{Hasenbusch:2001ne}
is also vital to enhance the efficiency of HMC.
In the context of lattice QCD with one-flavor, the basic idea of MP is to introduce 
an extra fermion flavor with mass $ m_h > m $, 
and rewrite the fermion determinant (\ref{eq:detD}) as 
\bea
\label{eq:MP}
\det H(m) &=& \det [H(m) H(m_h)^{-1}] \det H(m_h) \nn
          &=& \int d\phi^\dagger d\phi \exp\left( - \phi^\dagger H(m_h) H(m)^{-1} \phi \right)
              \int d\phi_h^\dagger d\phi_h \exp\left( - \phi_h^\dagger H(m_h)^{-1} \phi_h \right), \nn
&=& \int d\phi^\dagger d\phi \exp\left( - S_{pf}^{L} \right) 
    \int d\phi_h^\dagger d\phi_h \exp\left( - S_{pf}^H \right), 
\eea
where the dependence on the link variables $ U $ has been suppressed, 
$ S_{pf}^L = \phi^\dagger H(m_h) H(m)^{-1} \phi $, and 
$ S_{pf}^H = \phi_h^\dagger H(m_h)^{-1} \phi_h $.
This seemingly trivial modification turns out to have rather nontrivial consequences. 
First, the total number of CG iterations of computing the fermion forces 
$ -\partial S_{pf}^{L}/\partial A_l^a(\tau) $ and $ -\partial S_{pf}^{H}/\partial A_l^a(\tau) $ 
becomes less than that of computing the orginal fermion force $ -\partial S_{pf}/\partial A_l^a(\tau) $.
In other words, the HMC is speeded up by MP. Furthermore, $ \Delta\mathcal{H}$ may become smaller such that
the step-size ($\epsilon = 1/n $) of the integrator can be increased while maintaining the same  
acceptance rate. Thus the HMC efficiency (speed $\times$ acceptance rate) is 
enhanced by MP. Now it is straightforward to generalize MP
from one mass-preconditioner $ m_h $ 
to a cascade of mass-preconditioners $ m < m_{h_1} < m_{h_2} < \cdots < m_{h_N} $, 
which may lead to a higher efficiency for the HMC. Explicitly,  
\bea
\label{eq:MP_many}
\det H(m) &=& \det [H(m) H(m_{h_1})^{-1}] 
              \cdots \det [H(m_{h_{N-1}}) H(m_{h_{N}})^{-1}] \det [H(m_{h_{N}}) ]  \nn
&=& \int \prod_{i=0}^{N} d\phi_i^\dagger d\phi_i 
      \exp\left( -\phi_0^\dagger H(m_{h_1}) H(m)^{-1} \phi_0 - \cdots   \right. \nn
&&  \hspace{25mm}  \left.   -\phi_{N-1}^\dagger H(m_{h_N}) H(m_{h_{N-1}})^{-1} \phi_{N-1}   
                            -\phi_{N}^\dagger H(m_{h_N})^{-1} \phi_{N} \right). 
\eea
We refer Eqs. (\ref{eq:MP})-(\ref{eq:MP_many}) as the conventional (old) MP in lattice QCD.
     
In Ref. \cite{Chen:2014hyy}, 
we show that for lattice QCD with one-flavor domain-wall fermion (including all variants),
the fermion determinant can be written as a functional integral of an exact pseudofermion 
action with a positive-definite Hermitian operator $ H^{-1} $ 
(see Eq. (23) in Ref. \cite{Chen:2014hyy}), 
\bea
\label{eq:detDodwf}
  \frac{\det\Dodwf(m)}{\det\Dodwf(1)} 
= \frac{\det D_T(m)}{\det D_T(1)} 
= \int d \phi^\dagger d \phi \exp\left( - \phi^\dagger H^{-1}(m) \phi \right),
\eea
where
\bea
\label{eq:H}
H(m)^{-1} = P_{-} \left[I- k \Omega_{-}^T \frac{1}{H_T(m)} \Omega_{-} \right] P_{-} 
          + P_{+} \left[I+ k \Omega_{+}^T \frac{1}{H_T(1)- \Delta_+(m) P_+} \Omega_{+} \right] P_{+},  
\eea
$ P_{\pm} = (1 \pm \gamma_5)/2 $, $ H_T(m) = \gamma_5 R_5 D_T(m) $, $ R_5 $ is the reflection operator
in the fifth dimension, and $ D_T(m) $, $\Delta_{\pm}(m)$, and $ \Omega_\pm $ 
are defined by Eqs. (3), (15) and (18) in Ref. \cite{Chen:2014hyy}. 
We emphasize that the positive-definite Hermitian Dirac operator $ H(m)^{-1} $ is defined 
on the 4-dimensional lattice, while $ H_T(m) $ is a Hermitian Dirac operator 
defined on the 5-dimensional lattice.
In other words, in the EOFA, the DWF operator defined on the 5-dimensional lattice 
serves as a scaffold to give  
the positive-definite Hermitian Dirac operator $ H^{-1} $ on the 4-dimensional lattice 
such that $ \det H = \det D $, where $ D $ goes 
to the usual Dirac operator $ [ \gamma_\mu ( \partial_\mu + i A_\mu ) + m_q ]$ in the continuum limit. 
Note that in (\ref{eq:detDodwf})-(\ref{eq:H}), we have normalized the Pauli-Villars mass to one, and the 
quark mass to $ m = m_q/m_{PV} $, where $ m_q $ is the bare quark mass, and 
$ m_{PV} = 2 m_0 (1-d m_0) = 1/r $, as defined in Ref. \cite{Chen:2014hyy}.

A salient feature of the exact pseudofermion action for one-flavor DWF is 
that it can be decomposed into $\pm$ chiralities, as shown in (\ref{eq:H}). 
Thus the right-hand side of (\ref{eq:detDodwf}) can be rewritten as 
\bea
\int d\phi_-^\dagger d\phi_-  \exp\left( -\phi^\dagger P_- H^{-1}(m) P_- \phi \right) 
\int d\phi_+^\dagger d\phi_+  \exp\left( -\phi^\dagger P_+ H^{-1}(m) P_+ \phi \right), 
\eea
where
\bea
P_- \phi = 
\begin{pmatrix}
0 \\
\phi_- 
\end{pmatrix}, 
\hspace{4mm}    
P_+ \phi = 
\begin{pmatrix}
\phi_+ \\
0
\end{pmatrix}. 
\eea   
(Note that $ \phi_-/\phi_+ $ corresponds to $ \phi_1/\phi_2 $ in Ref. \cite{Chen:2014hyy}.) 
The pseudofermion actions of $\pm$ chiralites give two different fermion forces 
in the molecular dynamics of HMC.
In general, the fermion force coming from the pseudofermion action of $ \phi_- $ 
is much small than that of $ \phi_+ $.
Thus the gauge-momentum update by these two different fermion forces 
can be performed at two different time scales, according to the multiple-time scale method.    

Next, we consider the mass-preconditioning (MP) for the EOFA. 
According to (\ref{eq:MP}), (\ref{eq:MP_many}) and (\ref{eq:detDodwf}), MP for the EOFA can be written as  
\bea
\label{eq:EOFA_oldMP}
\frac{\det\Dodwf(m)}{\det\Dodwf(1)} = \frac{\det D_T(m)}{\det D_T(1)}  
= \frac{\det D_T(m)}{\det D_T(m_{h_1})} \frac{\det D_T(m_{h_1})}{\det D_T(m_{h_2})} 
    \cdots 
 \frac{\det D_T(m_{h_N})}{\det D_T(1)},  
\eea
where $ m_0 \equiv m < m_{h_1} < \cdots < m_{h_N} < 1 \equiv m_{N+1} $,  
\bea
\label{eq:pfij}
\frac{\det D_T(m_{i})}{\det D_T(m_{i+1})}
= \int d\phi_{i}^\dagger d\phi_{i} \exp\left( -\phi_i^\dagger H^{-1}(m_{i+1},m_{i}) \phi_i \right), 
\eea
\bea
\label{eq:Hij}
H^{-1}(m_j,m_i) 
&=& P_{-} \left[ I- k(m_j,m_i)\Omega_{-}^T \frac{1}{H_T(m_i)} \Omega_{-} \right] P_{-} \nn
&+& P_{+} \left[ I+ k(m_j,m_i) 
                \Omega_{+}^T \frac{1}{H_T(m_j)-\Delta_{+}(m_j,m_i) P_+} \Omega_{+} \right] P_{+},  
\eea
\bea
\label{eq:delta_ij}
\Delta_{\pm}(m_j,m_i) &\equiv& R_5\left\{ M_{\pm}(m_j)-M_{\pm}(m_i)\right\} 
    = k(m_j,m_i) \Omega_\pm \Omega^T_\pm,   \\
\label{eq:k_ij}
k(m_j,m_i) 
&\equiv& \frac{2c(m_j - m_i)}{(1+m_j-2cm_j\lambda)(1+m_i-2cm_i\lambda)}. 
\eea
Note that if setting $ m_j = 1 $ and $m_i = m $, then (\ref{eq:Hij}) reduces to (\ref{eq:H}), and  
(\ref{eq:delta_ij})-(\ref{eq:k_ij}) to Eqs. (16)-(17) in Ref. \cite{Chen:2014hyy}.

Now we call (\ref{eq:EOFA_oldMP}) the old MP for EOFA, which 
we have used in Refs. \cite{Chen:2014hyy,Chen:2014bbc}, with one heavy mass preconditioner.
In this paper, we introduce a new MP for the EOFA.

\section{Mass Preconditioning for the EOFA}

A vital observation for the mass preconditioning for EOFA is that the pseudofermion action 
of each chirality can be expressed as the ratio of two fermion determinants. 
This can be seen as follows.
\bea
\label{eq:det_minus}
\frac{\det D_T(m)}{\det D_T(m,1)}
&=& \int d\phi_-^\dagger d\phi_- \exp\left( -\phi^\dagger P_- H^{-1}(m) P_- \phi \right), \\
\label{eq:det_plus}
\frac{\det D_T(m,1)}{\det D_T(1)}
&=& \int d\phi_+^\dagger d\phi_+ \exp\left( -\phi^\dagger P_+ H^{-1}(m) P_+ \phi \right), 
\eea
where $ H(m)^{-1} $ is defined in (\ref{eq:H}), $ D_T(m_1,m_2) $ is defined by Eq.(11) 
in Ref. \cite{Chen:2014hyy}, 
\bea
\label{eq:DTm1m2}
D_T(m_1,m_2) \equiv 
\begin{pmatrix}
      W-m_0+M_+(m_1)      &  \sigma \cdot t \\
-(\sigma\cdot t)^{\dag} &   W-m_0+M_-(m_2)
\end{pmatrix},
\eea
and $ D_T(m) \equiv D_T(m,m) $.  
For consistency check, multiplying (\ref{eq:det_minus}) and (\ref{eq:det_plus}) on both hand sides
recovers (\ref{eq:detDodwf}), i.e., 
\bea
\label{eq:EOFA_noMP_D}
\frac{\det D_T(m)}{\det D_T(1)} = \frac{\det D_T(m)}{\det D_T(m,1)} \frac{\det D_T(m,1)}{\det D_T(1)}. 
\eea
The derivations of (\ref{eq:det_minus}) and (\ref{eq:det_plus}) are straightforward, 
similar to those leading to Eqs. (15) and (19) in Ref. \cite{Chen:2014hyy},   
using the Schur decompositions. 
 
Now, consider the old MP with one heavy mass preconditioner $ m_h $ ($ m < m_h < 1 $),  
(\ref{eq:EOFA_oldMP}) can be rewritten as 
\bea
\label{eq:MP_old}
\frac{\det D_T(m)}{\det D_T(1)} &=& \frac{\det D_T(m)}{\det D_T(m_h)} \frac{\det D_T(m_h)}{\det D_T(1)} \nn
&=& \frac{\det D_T(m)}{\det D_T(m,m_h)} \frac{\det D_T(m,m_h)}{\det D_T(m_h)} 
  \frac{\det D_T(m_h)}{\det D_T(m_h,1)} \frac{\det D_T(m_h,1)}{\det D_T(1)},   
\eea
resulting in four pseudofermion actions, each corresponds to one of the ratios of fermion determinants, 
with chiralities $-$, $+$, $-$, and $+$ respectively.  
Thus there are four different fermion forces, each corresponds to one of the pseudofermion actions, 
as shown in Fig. 2(a) of Ref. \cite{Chen:2014bbc}. 

Now we introduce the shorthand symbol $ (m_1, m_2) $ to denote $ \det D_T(m_1,m_2) $.
Thus we have  
\bea
\label{eq:m1m2_m3m4}
\frac{(m_3, m_4)}{(m_1, m_2)} \Leftrightarrow \frac{\det D_T(m_3,m_4)}{\det D_T(m_1, m_2)}.
\eea
Here we are only interested in two special cases: either $ m_1=m_3 $ or $ m_2=m_4 $.    
\bea
\label{eq:m1m2_m1m4}
\frac{(m_3, m_4)}{(m_3, m_2)} 
&\Leftrightarrow& \int d\phi_-^\dagger d\phi_- \exp \left( 
 -\phi^\dagger P_- \left[1-k(m_2,m_4) \Omega_-^T \frac{1}{H_T(m_3,m_4)} \Omega_- \right]P_- \phi \right), \\
\label{eq:m1m2_m3m2}
\frac{(m_3, m_4)}{(m_1, m_4)} 
&\Leftrightarrow& \int d\phi_+^\dagger d\phi_+ \exp \left(
-\phi^\dagger P_+ \left[ 1+k(m_1,m_3) \Omega_+^T \frac{1}{H_T(m_3,m_4)} \Omega_+ \right] P_+ \phi\right),   
\eea
where $ k(m_j,m_i) $ is defined in (\ref{eq:k_ij}), and $ H_T(m_j,m_i) = \gamma_5 R_5 D_T(m_j,m_i) $.
Note that for $ m_1 = m_3 $ (i.e., the masses on the left column are equal),
(\ref{eq:m1m2_m3m4}) becomes the pseudofermion action with negative chirality (\ref{eq:m1m2_m1m4}); 
while for $ m_2 = m_4 $ (the masses on the right column are equal), it becomes  
the pseudofermion action with positive chirality (\ref{eq:m1m2_m3m2}).
In the following, we will use the shorthand symbols (\ref{eq:m1m2_m1m4}) and (\ref{eq:m1m2_m3m2}) 
to refer to the fermion determinant together with the pseudofermion fermion action 
with $\pm$ chirality.

In the following, for generality, 
we consider $ (m,m)/(M,M) $ instead of $ (m,m)/(1,1) $, where $ m < M < 1 $. 
With the shorthand symbol, (\ref{eq:EOFA_noMP_D}) is re-written as 
\bea
\label{eq:noMP}
\frac{(m,m)}{(M,M)} = \frac{(m,m)}{(m,M)} \frac{(m,M)}{(M,M)},  \hspace{4mm} m < M < 1,  
\eea
which gives two pseudofermion actions with chiralities $\{ -, + \}$ respectively. 
Alternatively, we can write (\ref{eq:noMP}) as 
\bea
\label{eq:noMP_P}
\frac{(m,m)}{(M,M)} = \frac{(m,m)}{(M,m)} \frac{(M,m)}{(M,M)},   
\eea
which gives two pseudofermion actions with chiralities $\{ +, - \}$ respectively. 
Note that in Ref. \cite{Chen:2014hyy}, 
we have only presented (\ref{eq:noMP}), but omitted (\ref{eq:noMP_P}).   
In fact, (\ref{eq:noMP}) and (\ref{eq:noMP_P}) are related by the ``parity" operation,  
$ {\cal P} $, which is defined as  
swapping the masses of the left and right columns in the shorthand symbol, i.e., 
\BAN
\label{eq:parity}
{\cal P} \frac{(m_3, m_4)}{(m_1, m_2)} = \frac{(m_4, m_3)}{(m_2, m_1)}, \hspace{4mm} {\cal P}^2 = \Id. 
\EAN
Thus 
\BAN
{\cal P} \frac{(m_3, m_4)}{(m_3, m_2)} = \frac{(m_4, m_3)}{(m_2, m_3)},  \hspace{4mm}
{\cal P} \frac{(m_3, m_4)}{(m_1, m_4)} = \frac{(m_4, m_3)}{(m_4, m_1)}.  
\EAN
In short, the parity operation changes a pseudofermion action with positive chirality to 
the corresponding one with negative chirality, and vice versa. 
Under parity, (\ref{eq:noMP}) becomes (\ref{eq:noMP_P}), and vice versa, thus they are called 
parity partners.
Presumably, parity partners have compatible HMC efficiencies. 
However, in practice, one of them may turn out to be slightly better than the other.  
In the following, we only write down one of the parity partners.  

Besides (\ref{eq:m1m2_m1m4}) and (\ref{eq:m1m2_m3m2}), 
we also introduce the shorthand symbols $ F_-(m_3; m_4, m_2)$ and $ F_+(m_3, m_1; m_4)$  
to denote the fermion forces corresponding to 
$(m_3, m_4)/(m_3, m_2)$ and $(m_3, m_4)/(m_1, m_4)$ respectively. That is
\BAN
F_-(m_3; m_4, m_2) \Leftrightarrow F_{-} \left[ \frac{(m_3,m_4)}{(m_3,m_2)} \right], \hspace{4mm}
F_+(m_3, m_1; m_4) \Leftrightarrow F_{+} \left[ \frac{(m_3,m_4)}{(m_1,m_4)} \right].
\EAN
Similarly, we use $ N^{cg}_-(m_3; m_4, m_2) $ and $ N^{cg}_+(m_3, m_1; m_4) $     
to denote the total number of CG iterations (per one trajectory) in computing the fermion 
forces $ F_-(m_3; m_4, m_2)$ and $ F_+(m_3, m_1; m_4)$ respectively,   
together with that in generating the corresponding $ \phi_\pm $ from the Gaussian noises 
in the beginnning of the trajectory. In other words, $ N^{cg}_\pm $ counts all CG iterations 
in one HMC trajectory, and it always refers to the averaged value  
over a large number of HMC trajectories after thermalization. 

In general, for (\ref{eq:noMP}), the fermion forces satisfy the inequality
\BAN
\label{eq:F_noMP}
F_- (m;m,M) < F_+ (m,M;M),  
\EAN
thus they are amenable to the multiple-time scale method. 
Here the magnitiude of the fermion force always refers to its average over all link variables, i.e.,  
\BAN
F_{\pm} = \frac{1}{4 N_l^3 N_t} \left[ \sum_{x,\mu} \sum_{a=1}^8 \left(F^a_\mu(x)\right)^2 \right]^{1/2}. 
\EAN
where $ N_l $ and $ N_t $ denote the number of sites in the spatial and time directions. 

For MP with one heavy mass preconditioner, (\ref{eq:MP_old}) can be rewritten as 
\bea
\label{eq:old_MP}
\frac{(m,m)}{(M,M)} = \frac{(m,m)}{(m_h,m_h)} \frac{(m_h,m_h)}{(M,M)} 
= \frac{(m,m)}{(m,m_h)} \frac{(m,m_h)}{(m_h,m_h)} \frac{(m_h,m_h)}{(m_h,M)} \frac{(m_h,M)}{(M,M)},   
\eea 
which gives four pseudofermion actions with chiralities $\{ -, +, -, + \} $ respectively.
In general, the fermion forces are ordered according to 
\BAN
\label{F_MP_old}
  F_- (m;m,m_h) < F_+ (m,m_h;m_h) < F_- (m_h;m_h,M) < F_+ (m_h,M;M),   
\EAN
thus they are amenable to the multiple-time scale (MTS) method. 
Now if we put all fermion forces at the same level of MTS, 
we find that using MP with one heavy mass preconditioner (\ref{eq:old_MP}) takes less 
CG iterations than that without MP (\ref{eq:noMP}), i.e.,  
\BAN
&& N^{cg}_- (m;m,m_h) + N^{cg}_+ (m,m_h; m_h) + N^{cg}_- (m_h; m_h,M) + N^{cg}_+ (m_h,M;M) \nn
< && N^{cg}_- (m;m,M) + N^{cg}_+ (m,M;M). 
\EAN 
If MTS is also turned on, the gain of using MP is even larger. 
Note that (\ref{eq:old_MP}) is only one of the 4 parity partners, i.e., 
\BAN
& & \frac{(m,m)}{(m,m_h)} \frac{(m,m_h)}{(m_h,m_h)} \frac{(m_h,m_h)}{(m_h,M)} \frac{(m_h,M)}{(M,M)}   
  = \frac{(m,m)}{(m_h,m)} \frac{(m_h,m)}{(m_h,m_h)} \frac{(m_h,m_h)}{(m_h,M)} \frac{(m_h,M)}{(M,M)} \\  
&=& \frac{(m,m)}{(m,m_h)} \frac{(m,m_h)}{(m_h,m_h)} \frac{(m_h,m_h)}{(M,m_h)} \frac{(M,m_h)}{(M,M)}   
  = \frac{(m,m)}{(m_h,m)} \frac{(m_h,m)}{(m_h,m_h)} \frac{(m_h,m_h)}{(M,m_h)} \frac{(M,m_h)}{(M,M)},  
\EAN 
which give pseudofermion actions with chiralities 
$\{-,+,-,+\} $, $\{+,-,-,+\} $, $\{-,+,+,-\} $, and $\{+,-,+,-\} $ respectively.
Presumably, parity partners have compatible HMC efficiencies. 
However, in practice, one of them may turn out to be slightly better than the others.  
In the following, we only write down one of the parity partners.

It is straightforward to generalize MP from one heavy mass 
to a cascade of heavy masses ($ m < m_{h_1} < m_{h_2} < \cdots < m_{h_N} < M $), 
which may lead to even higher efficiency for HMC. 
Explicitly, we have 
\bea
\label{eq:old_MP_N}
\frac{(m,m)}{(M,M)} &=& \frac{(m,m)}{(m_{h_1},m_{h_1})} \frac{(m_{h_1},m_{h_1})}{(m_{h_2},m_{h_2})} 
                      \cdots \frac{(m_{h_N},m_{h_N})}{(M,M)}  \nn
&=& \frac{(m,m)}{(m,m_{h_1})} \frac{(m,m_{h_1})}{(m_{h_1},m_{h_1})} 
    \frac{(m_{h_1},m_{h_1})}{(m_{h_1},m_{h_2})} \frac{(m_{h_1},m_{h_2})}{(m_{h_2},m_{h_2})} 
    \cdots \frac{(m_{h_N},m_{h_N})}{(m_{h_N},M)} \frac{(m_{h_N},M)}{(M,M)}.   
\eea
Now (\ref{eq:old_MP_N}) seems to cover all MP schemes (plus their parity partners) for the EOPA.  
Nevertheless, due to the chiral structure of the exact one-flavor pseudofermion action, 
there exists a new MP scheme for the EOFA. 

First we consider the MP of $ (m,m)/(M,M) $ with one heavy mass preconditioner 
$ m_h $ ($ m < m_h < M $). We observe that it is possible to write
\bea
\label{eq:new_MP}
\frac{(m,m)}{(M,M)} = \frac{(m,m)}{(m,m_h)} \frac{(m,m_h)}{(M,m_h)} \frac{(M,m_h)}{(M,M)},  
\eea
which gives three pseudofermion actions with chiralities $\{ -, +, - \}$ respectively.
This is a new MP scheme, different from the old MP (\ref{eq:old_MP}) with 4 pseudofermion actions. 
At first sight, it is unclear whether the new MP (\ref{eq:new_MP}) 
is more efficient than the old one (\ref{eq:old_MP}). 
Nevertheless, it turns out that this is the case, and the following inequality 
holds for all cases we have studied, with or without MTS. 
\BAN
N^{cg}_+ (m,M;m_h) <  N^{cg}_+ (m,m_h;m_h) + N^{cg}_- (m_h;m_h,M). 
\EAN
Moreover, using the new MP scheme also yields a smaller $ \Delta {\cal H} $ 
and higher acceptance rate than the old MP. 
It is straightforward to generalize the new MP scheme from one heavy mass preconditioner (\ref{eq:new_MP}) 
to a cascade of heavy mass preconditioners ($ m < m_{h_1} < m_{h_2} < \cdots < m_{h_N} < M $),  
\bea
\label{eq:new_MP_N}
%
\frac{(m,m)}{(M,M)} &=& \frac{(m,m)}{(m,m_{h_1})} 
                        \frac{(m,m_{h_1})}{(m_{h_2},m_{h_1})} 
                        \frac{(m_{h_2},m_{h_1})}{(m_{h_2},m_{h_3})} 
\cdots
\frac{(m_{h_{N-1}},m_{h_N})}{(M,m_{h_N})} 
\frac{(M,m_{h_N})}{(M,M)}.  
\eea
Obviously, (\ref{eq:new_MP_N}) has only one parity partner. 
Equations (\ref{eq:new_MP}) and (\ref{eq:new_MP_N}) (plus their parity partners) 
are the main results of this paper.
In the next section, we compare the HMC efficiency with the new MP to those with   
the old MP and without MP, for $ N_f = 1 $ and $ N_f = 1+1+1+1 $ QCD 
with the optimal domain-wall quarks, on the $ 24^3 \times 48 $ lattice.

\section{Numerical Tests}  

\subsection{$ N_f = 1 $ QCD}

\begin{figure*}[tb]
\begin{center}
\begin{tabular}{@{}c@{}c@{}c@{}}
\includegraphics*[width=6cm,clip=true]{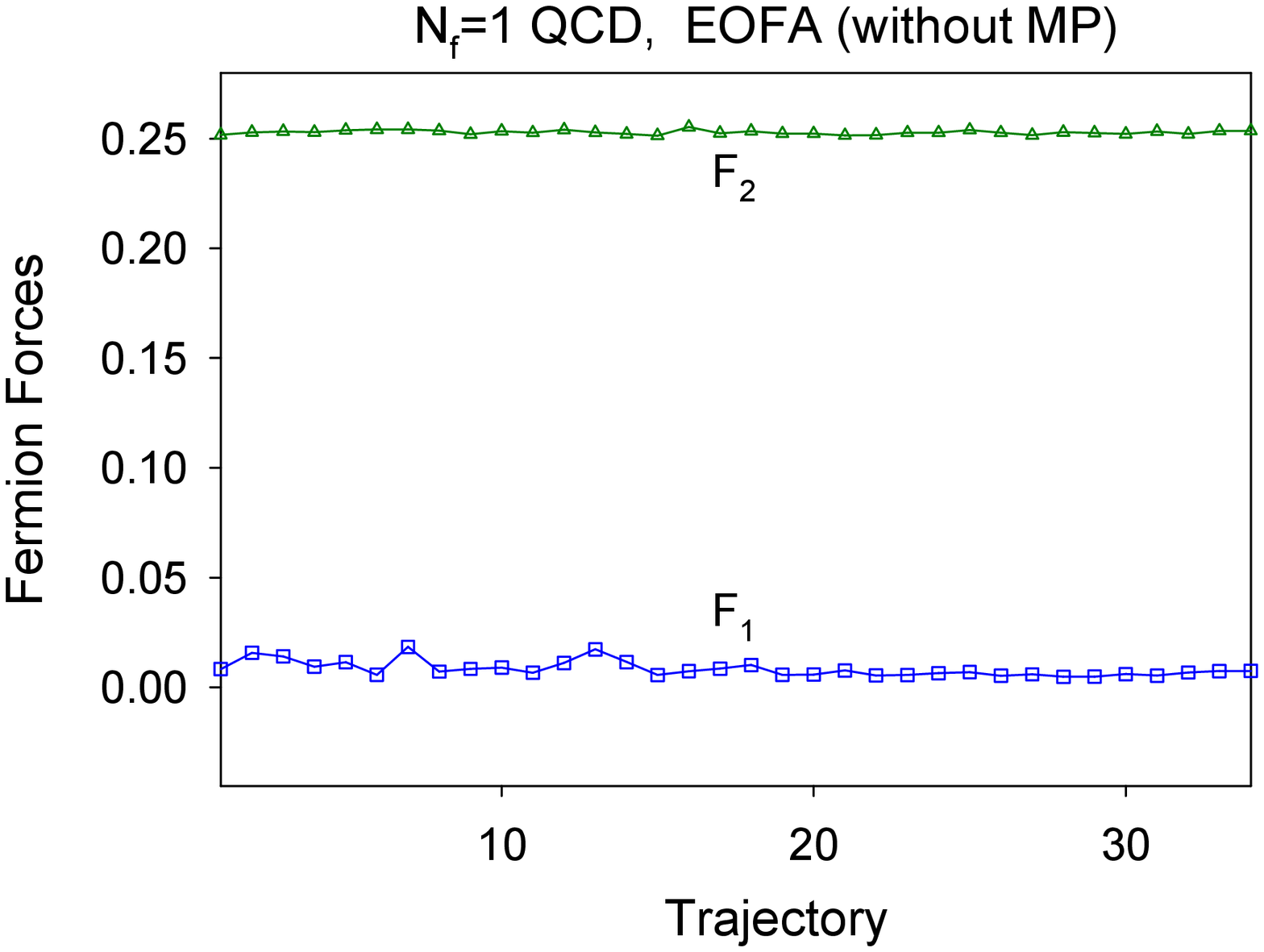}
&
\includegraphics*[width=5.1cm,clip=true]{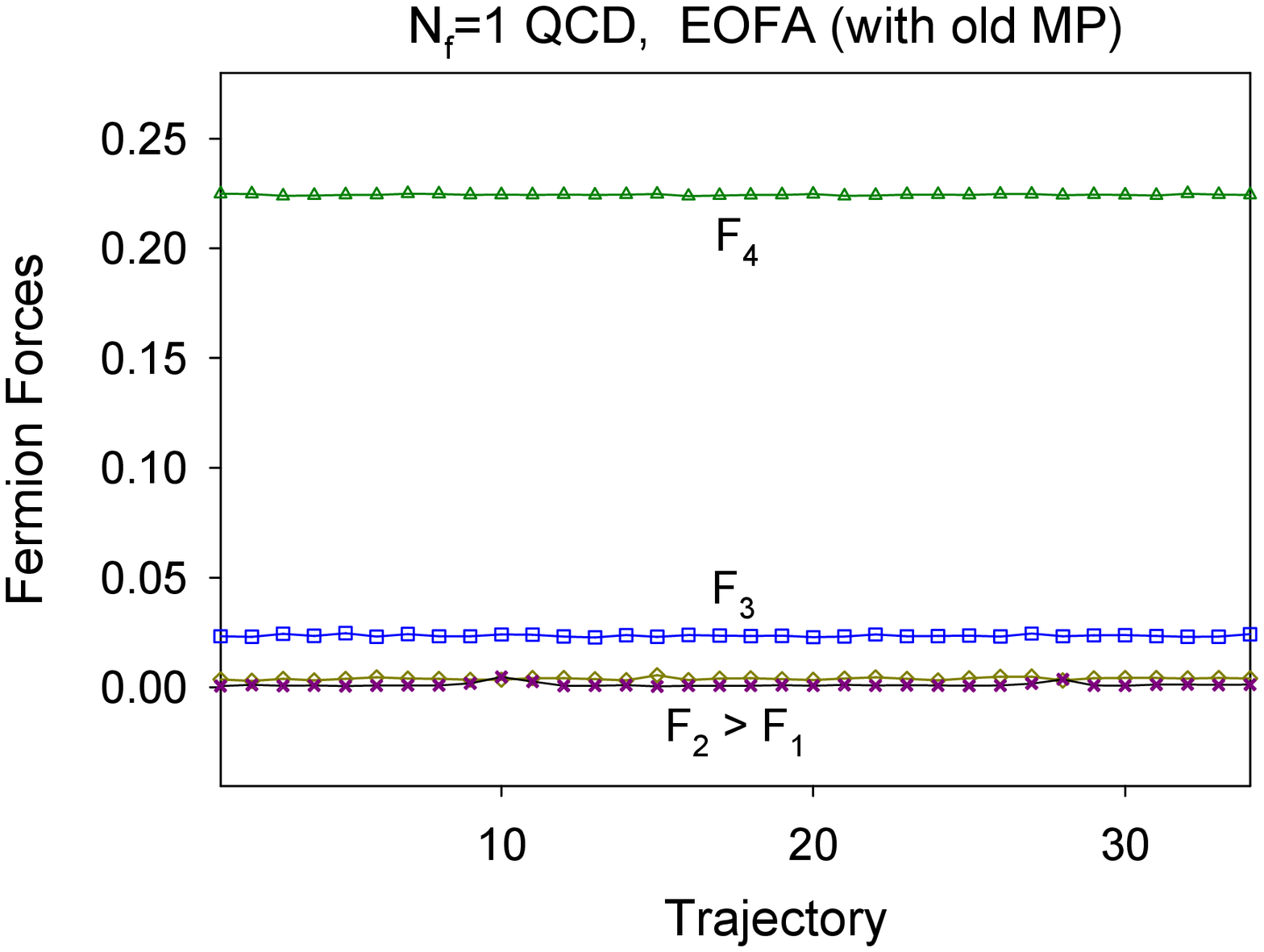}
&
\includegraphics*[width=5.1cm,clip=true]{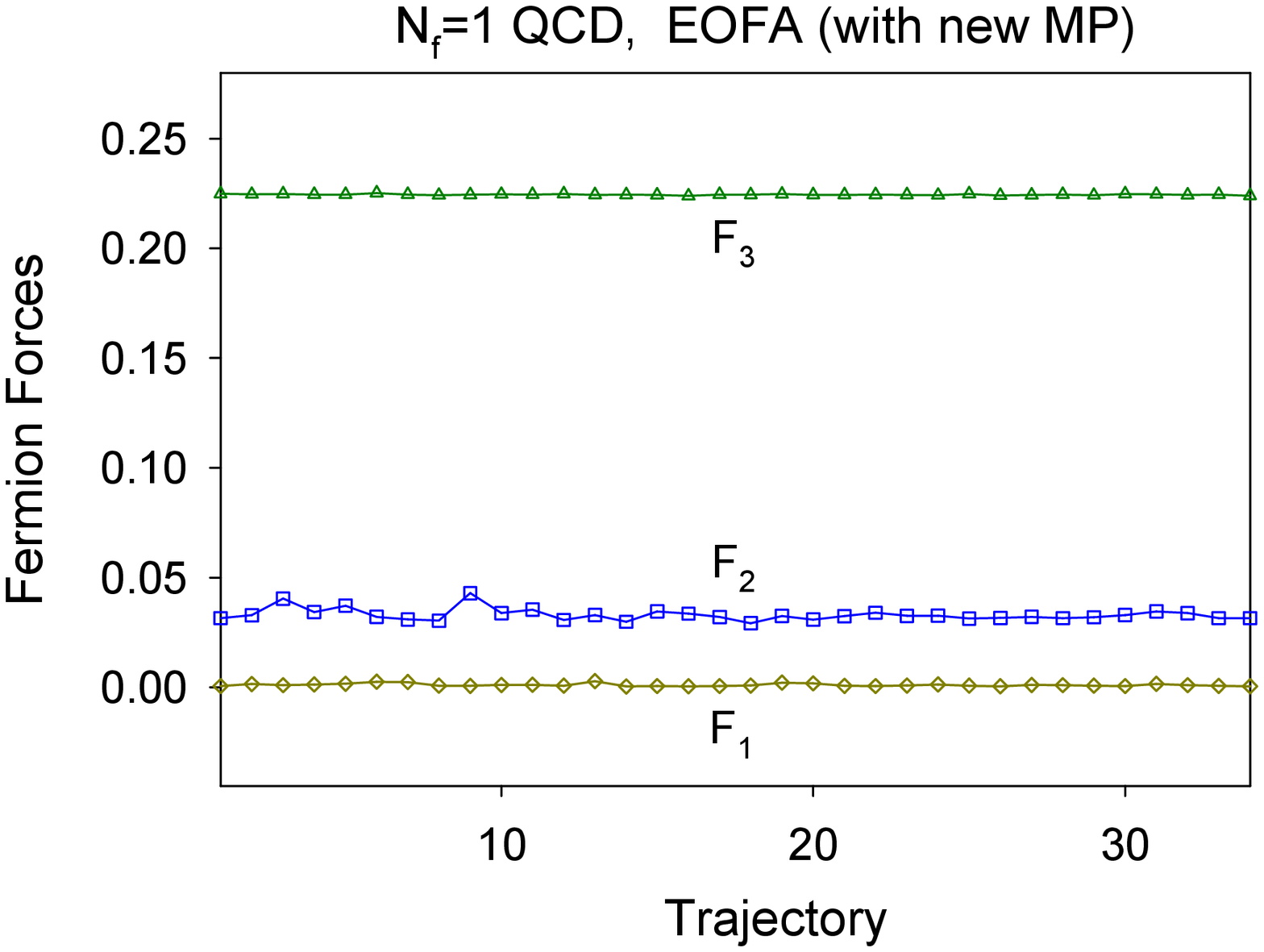}
\end{tabular}
\caption{The maximum forces of the fermion fields  
         versus the trajectory in the HMC of one-flavor QCD with the optimal DWF, 
         for 3 different MP schemes applying to the EOFA of $(m_q,m_q)/(m_{PV},m_{PV})$.}
\label{fig:Fforces_Nf1}
\end{center}
\end{figure*}

Using Nvidia GTX-970 (4 GB device memory), 
we perform the HMC of one-flavor QCD on the $ 24^3 \times 48 $ lattice,  
with the optimal domain-wall quark ($ N_s = 16 $, $ m_0 = 1.3 $,  
$ \lambda_{max}/\lambda_{min} = 6.20/0.05 $),  
and the Wilson plaquette gauge action at $ \beta = 6.20 $. 
The optimal weights $ \{ \omega_s \} $ with $ R_5 $ symmetry are computed 
with the Eq. (9) in Ref. \cite{Chiu:2015sea}.
The bare quark mass is $ m_q = 0.005 $, and the mass of the heavy mass-preconditioner 
$ m_h = 0.1 $. 
In the following, we will use the bare quark masses 
in the shorthand symbol (\ref{eq:m1m2_m3m4}), and it is understood that 
they are normalized by the Pauli-Villars mass $ m_{PV} = 2 m_0 = 2.6 $.
Then, with the shorthand symbol (\ref{eq:m1m2_m3m4}),   
the 3 different MP schemes read:
\bea
\label{eq:Nf1}
\frac{(m_q,m_q)}{(m_{PV},m_{PV})} &=& \frac{(m_q,m_q)}{(m_q,m_{PV})} \frac{(m_q,m_{PV})}{(m_{PV},m_{PV})},
\hspace{2mm} \textmd{(without MP)} \\  
\label{eq:Nf1_newMP}
\frac{(m_q,m_q)}{(m_{PV},m_{PV})} 
&=& \frac{(m_q,m_q)}{(m_q,m_h)} \frac{(m_q,m_h)}{(m_{PV},m_h)} \frac{(m_{PV},m_h)}{(m_{PV},m_{PV})},  
\hspace{2mm} \textmd{(the new MP)} \\  
\label{eq:Nf1_oldMP}
\frac{(m_q,m_q)}{(m_{PV},m_{PV})} 
&=& \frac{(m_q,m_q)}{(m_q,m_h)} \frac{(m_q,m_h)}{(m_h,m_h)} 
    \frac{(m_h,m_h)}{(m_h,m_{PV})} \frac{(m_h,m_{PV})}{(m_{PV},m_{PV})},  
\hspace{2mm} \textmd{(the old MP)}.   
\eea 
Each factor on the RHS of (\ref{eq:Nf1})-(\ref{eq:Nf1_newMP}) 
can be written as a functional integral with the pseudofermion action of negative chirality 
(\ref{eq:m1m2_m1m4}) or positive chirality (\ref{eq:m1m2_m3m2}).
The fermion forces coming from the first and the second factor on the RHS of (\ref{eq:Nf1}) 
are denoted by $ F_{-}(m_q;m_q,m_{PV}) \equiv F_{1}^{2f}$ 
and $ F_{+}(m_q,m_{PV};m_{PV}) \equiv F_2^{2f}$ respectively, where the superscript 
$2f$ stands for the two factors on the RHS of (\ref{eq:Nf1}).
Similarly, for the old MP (\ref{eq:Nf1_oldMP}), the fermion forces are denoted by 
$F_{1}^{4f}$, $F_2^{4f}$, $F_3^{4f}$, and $F_4^{4f}$, in the same order 
as the RHS of (\ref{eq:Nf1_oldMP}). Finally, for the new MP (\ref{eq:Nf1_newMP}), 
the fermion forces are denoted by $F_{1}^{3f}$, $F_2^{3f}$, and $F_3^{3f}$, in the same order 
as the RHS of (\ref{eq:Nf1_newMP}).
  
In the molecular dynamics, we use the Omelyan integrator \cite{Omelyan:2001} 
and the multiple-time scale method \cite{Sexton:1992nu}, for 3 different MP schemes.  
Starting with the same initial thermalized configuration, 33 HMC trajectories are generated for each case. 

In Fig. \ref{fig:Fforces_Nf1}, we plot the maximum fermion forces (averaged over all links)
among all momentum updates in each trajectory, for 3 different MP schemes. 
With the length of the HMC trajectory equal to one, we set three different time scales, namely, 
$ (k_0, k_1, k_2 ) $, where the smallest time step (for the link update) 
in the molecular dynamics is $1/(2k_0)$. 
The fields are updated according to the following assignment:
\BAN
&&k_0 : U_{\mu} (\textmd{gauge field}), \\
&&k_1 : \phi_2^{2f}, \phi_3^{3f}, \phi_4^{4f}, \\
&&k_2 : \phi_1^{2f}, \phi_1^{3f}, \phi_2^{3f}, \phi_1^{4f}, \phi_2^{4f}, \phi_3^{4f}, 
\EAN
where the superscripts $2f$, $3f$, and $4f$ refer to the number of factors on 
the RHS of (\ref{eq:Nf1}), (\ref{eq:Nf1_newMP}), and (\ref{eq:Nf1_oldMP}) respectively, 
and the subscripts refer to which factor on the RHS of (\ref{eq:Nf1})-(\ref{eq:Nf1_oldMP}).
For example, $\phi_2^{4f}$ denotes the pseudofermion field corresponding to 
the second factor $ (m_q,m_h)/(m_h,m_h) $ in (\ref{eq:Nf1_oldMP}). 
In our simulations, we set $ (k_0, k_1, k_2) = (480, 12, 6) $, 
then the number of link updates is $ 2 \times 480 = 960 $,   
and the numbers of momentum updates for $(k_0, k_1, k_2)$ are $ (961, 25, 13) $ respectively.
The gauge forces are not plotted in Fig. \ref{fig:Fforces_Nf1}, 
with the averaged values: 5.7345(4) (without MP), 5.3756(3) (old MP), and 5.3761(2) (new MP). 
   
\begin{figure*}[tb]
\begin{center}
\includegraphics*[width=10cm,clip=true]{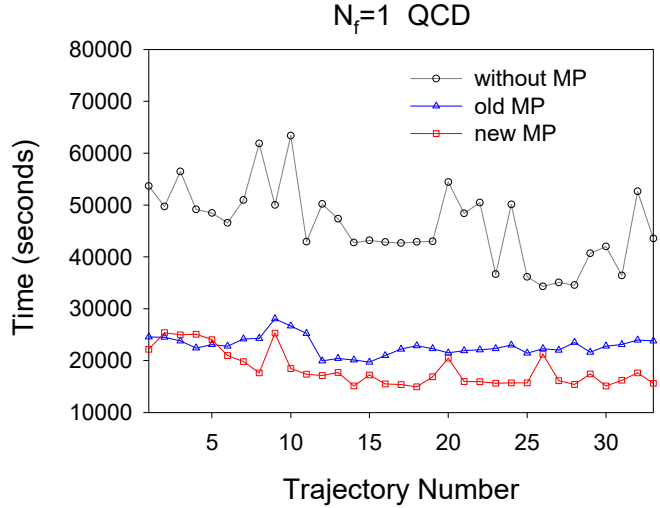}
\caption{The elapsed time versus the trajectory in the HMC of one-flavor QCD with the optimal DWF, 
         for 3 different MP schemes applying to the EOFA of $(m_q,m_q)/(m_{PV},m_{PV})$.}
\label{fig:Time_Nf1}
\end{center}
\end{figure*}

In Fig. \ref{fig:Time_Nf1}, we plot the elapsed time versus the HMC trajectory, for  
3 different MP schemes. The statistics of the elapsed time, the acceptance rate, 
and the maximum fermion forces are summarized as follows.
\begin{center}
\begin{tabular}{ccccccc}
\hline
  & $ F_{\phi_1} $ & $ F_{\phi_2} $ & $ F_{\phi_3} $ & $ F_{\phi_4} $ & Time/traj.(secs) & Acceptance  \\
\hline
without MP & 0.0830(6) & 0.2529(2) &           &           & 46162(1287) & 0.97(3)  \\ 
old MP     & 0.0012(1) & 0.0040(1) & 0.0234(1) & 0.2244(1) & 22839(315)  & 0.88(6)   \\
new MP     & 0.0011(1) & 0.0348(5) & 0.2243(1) &           & 18346(594)  & 0.88(6)  \\ 
\hline
\end{tabular}
\end{center}
From the data above, the HMC speed with the new MP is $\sim 2.5 $ times of that without MP, 
and $ \sim 1.25 $ times of that with the old MP. 
If the acceptance rate is also taken into account, 
the HMC efficiency (speed $\times$ acceptance rate)
with the new MP is about $\sim 120\%$ higher that without MP, 
and $\sim 25\%$ higher than that with the old MP.

\subsection{$ N_f = 1+1+1+1 $ QCD}

Since all physical quark masses are non-degenerate, lattice studies are required to simulate 
$ N_f = 1+1+1+1+1 $ QCD with $ (u,d,s,c,b) $ quarks. However, for domain-wall fermions, 
to simulate $ N_f = 1+1 $ amounts to simulate $ N_f = 2+1 $, as pointed out in \cite{Chen:2017kxr}. 
Similarly, to simulate $ N_f = 1+1+1+1 $ amounts to simulate $ N_f = 2+2+1+1 $, i.e.,  
\bea
&& \frac{\det \Dodwf(m_u)}{\det \Dodwf(m_{PV})}   
   \frac{\det \Dodwf(m_d)}{\det \Dodwf(m_{PV})}   
   \frac{\det \Dodwf(m_s)}{\det \Dodwf(m_{PV})}   
   \frac{\det \Dodwf(m_c)}{\det \Dodwf(m_{PV})}    \nn 
&=& 
\left( \frac{\det \Dodwf(m_s)}{\det \Dodwf(m_{PV})} \right)^2  
\left( \frac{\det \Dodwf(m_c)}{\det \Dodwf(m_{PV})} \right)^2  
\frac{\det \Dodwf(m_d)}{\det \Dodwf(m_{c})}       
\frac{\det \Dodwf(m_u)}{\det \Dodwf(m_{s})},
\label{eq:Nf4}       
\eea
where only one of the 12 different ways of writing the expression of $ N_f = 2+2+1+1$ is given.   
Obviously, it is better to simulate $ N_f = 2+2+1+1 $ than $ N_f = 1+1+1+1 $, 
since the simulaton of 2-flavors is most likely faster than the simulaton of one-flavor.
In the following, it is understood that the simulation of $ N_f = 1+1+1+1 $ QCD is 
performed by simulating the equivalent $N_f=2+2+1+1$ QCD, according to (\ref{eq:Nf4}).      

Using Nvidia GTX-1060 (6 GB device memory), 
we perform the HMC of $N_f=1+1+1+1 $ QCD on the $ 24^3 \times 48 $ lattice,  
with the optimal domain-wall quarks ($ N_s = 16 $, $ m_0 = 1.3 $,  
$ \lambda_{max}/\lambda_{min} = 6.20/0.05 $),  
and the Wilson plaquette gauge action at $ \beta = 6.20 $. 
The optimal weights $ \{ \omega_s \} $ for the 2-flavors action are computed with the Eq. (12) 
in \cite{Chiu:2002ir}, while those with $ R_5 $ symmetry for the EOFA are computed 
with the Eq. (9) in Ref. \cite{Chiu:2015sea}.
The bare quark masses are: $m_u = 0.005$, $m_d = 0.01 $, $m_s = 0.04$, and $m_c = 0.55$, 
where $m_s$ and $m_c$ are close to the physical bare quark masses.  

To simplify the test,  
only the EOFA of $ \det \Dodwf(m_d)/\det \Dodwf(m_{c})$ in (\ref{eq:Nf4}) 
are tested for 3 different MP schemes (old/new/none), with one mass preconditioner $ m_h = 0.1 $, 
while the EOFA of $ \det \Dodwf(m_u)/\det \Dodwf(m_{s})$ is simulated without MP.
For the simulation of 2-flavors determinants $ (\det \Dodwf(m_s)/\det \Dodwf(m_{PV}))^2 $ and 
$ (\det \Dodwf(m_c)/\det \Dodwf(m_{PV}))^2 $, the details have been presented in Ref. \cite{Chiu:2013aaa}.
Here MP is only applied to $ (\det \Dodwf(m_s)/\det \Dodwf(m_{PV}))^2 $ with 
one heavy mass preconditoner $ m_H = 0.4 $, while 
$ ( \det \Dodwf(m_c)/\det \Dodwf(m_{PV}) )^2 $ is simulated without MP. 
Setting the length of the HMC trajectory equal to one, we use five different time scales in MTS, 
namely, $ (k_0, k_1, k_2, k_3, k_4 ) $, where the smallest time step (for the link update) 
in the molecular dynamics is $1/(2k_0)$. The fields are updated according to the following assignment:
\BAN
&&k_0 : U_{\mu} (\textmd{gauge field}), \\
&&k_1 : \Phi^c, \Phi^s_{H} \\
&&k_2 : \Phi^s_{L}  \\
&&k_3 : \{\phi_1^{d}, \phi_2^{d}\}_{\textmd{(without MP)}},  
        \{\phi_2^d, \phi_3^d, \phi_4^d \}_{\textmd{(old MP)}}, 
        \{\phi_2^d, \phi_3^d \}_{\textmd{(new MP)}}, \\
&&k_4 : \{\phi_1^{u}, \phi_2^{u}\}, \{\phi_1^d \}_{\textmd{(new MP)}}, \{\phi_1^d \}_{\textmd{(old MP)}}. 
\EAN
Here $ \{ \Phi^c, \Phi^s_H, \Phi^s_L \}$ are the pseudofermion fields in the  
2-flavors actions corresponding to  
$ (\det \Dodwf(m_c)/\det \Dodwf(m_{PV}))^2 $, $ (\det \Dodwf(m_H)/\det \Dodwf(m_{PV}))^2 $, 
and $ (\det \Dodwf(m_s)/\det \Dodwf(m_H))^2 $ respectively. 
For the EOFA involving the $ \u $ quark (without MP),    
$ \{ \phi_1^{u}, \phi_2^{u}\} $ are the pseudofermion fields corresponding to 
$ \{ (m_u, m_u)/(m_u, m_s), (m_u, m_s)/(m_s, m_s) \} $. 
For the EOFA involving the $ \d $ quark, 
$ \{\phi_1^{d}, \phi_2^{d}\}_{\textmd{(without MP)}} $,  
$ \{\phi_1^d, \phi_2^d, \phi_3^d \}_{\textmd{(new MP)}} $,  
and $ \{\phi_1^d, \phi_2^d, \phi_3^d, \phi_4^d \}_{\textmd{(old MP)}} $    
are the pseudofermion fields 
corresponding to $ \{ (m_d, m_d)/(m_d, m_c), (m_d, m_c)/(m_c, m_c) \} $ (without MP),  
$ \{ (m_d, m_d)/(m_d, m_h), (m_d, m_h)/(m_c, m_h), (m_c, m_h)/(m_c, m_c) \} $ (the new MP), and 
$ \{(m_d, m_d)/(m_d, m_h), (m_d, m_h)/(m_h, m_h), (m_h, m_h)/(m_h, m_c), (m_h, m_c)/(m_c, m_c) \}$
(the old MP) respectively.    
In our simulations, we set $ (k_0, k_1, k_2, k_3, k_4) = (480, 48, 24, 12, 6) $,  
then the number of link updates is $ 2 \times 480 = 960 $,   
and the numbers of momentum updates for $(k_0, k_1, k_2, k_3, k_4)$ 
are $ (961, 97, 49, 25, 13) $, for each trajectory. 

\begin{figure*}[tb]
\begin{center}
\includegraphics*[width=10cm,clip=true]{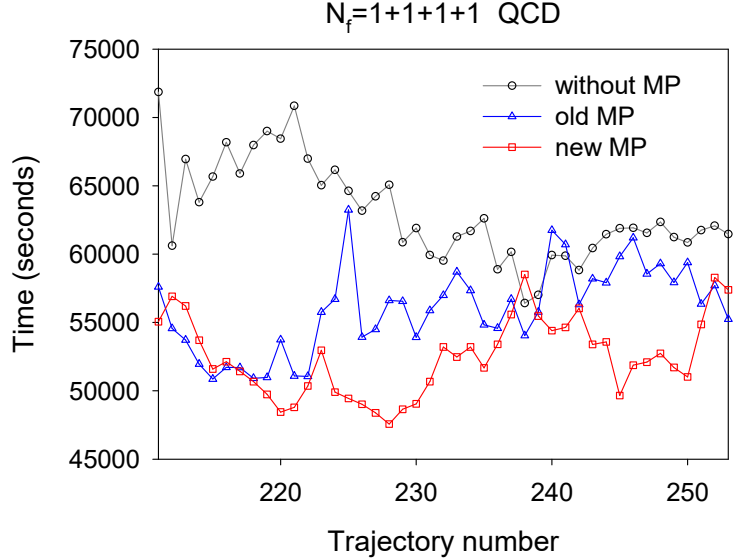}
\caption{The total elapsed time versus the trajectory in the HMC of $N_f=1+1+1+1 $ QCD with the optimal DWF, 
         for 3 different MP schemes applying to the EOFA of $ (m_d, m_d)/(m_c, m_c) $.} 
\label{fig:Time_Nf4}
\end{center}
\end{figure*}

Starting with the same initial thermalized configuration, 3 independent HMC simulations are performed 
with 3 different MP schemes for the EOFA of $ (m_d, m_d)/(m_c, m_c) $,   
and 44 HMC trajectories are generated for each case. 
In Fig. \ref{fig:Time_Nf4}, we plot the total elapsed time versus the HMC trajectory, 
for 3 different MP schemes. 
The statistics of 44 trajectories are listed below, for the time used in the simulation of EOFA 
of $ (m_d, m_d)/(m_c, m_c) $ (the 2nd column), the total elapsed time (the 3rd column), 
and the acceptance rate (the 4th column).
\begin{center}
\begin{tabular}{cccc}
\hline
       &  Time[$(m_d, m_d)/(m_c, m_c)$]/traj.(sec.)  &  Total time/traj.(sec.)  & Acceptance rate  \\
\hline
without MP & 29677(187)  & 60720(652)  & 0.86(5)  \\ 
old MP     & 26545(254)  & 56031(475)  & 0.80(6)  \\
new MP     & 21879(222)  & 52658(469)  & 0.91(4)  \\ 
\hline
\end{tabular}
\end{center}
From the second column, for the simulation of the EOFA of $ (m_d, m_d)/(m_c, m_c) $ only, 
the speed with the new MP is $\sim 1.36 $ times of that without MP, 
and $ \sim 1.21 $ times of that with the old MP. 
In other words, the new MP is about $21\%$ faster than the old MP.  
Since the simulation of $ (m_d, m_d)/(m_c, m_c) $ only constitutes about 40-50\% of the entire
HMC simulation, the total simulation time of $ N_f = 1+1+1+1 $ (in the third column) shows that
the new MP is only 6.4\% faster than the old MP.        
However, if the acceptance rate is also taken into account,  
the HMC efficiency (speed $\times$ acceptance rate) with the new MP 
is $ \sim 21\% $ higher than that with the old MP. 

Finally, we note that in our numerical tests of $N_f=1$ and $ N_f=1+1+1+1 $ QCD, we have not explored 
further enhancement of the new MP with a cascade of mass preconditioners, 
which is beyond the scope of this paper.    

\section{Concluding remarks}

Due to the chiral structure of the EOFA, there exists a novel mass preconditioning  
which only involves 3 chiral pseudofermion actions (\ref{eq:new_MP}) (or its parity partner),  
rather than the old MP which involves 4 chiral pseudofermion actions (\ref{eq:old_MP}) 
(or any one of its 3 parity partners), for MP with one heavy mass preconditioner. 
This can be generalized to a cascade of $ N $ heavy mass preconditioners, in which 
the new MP only involves $ N+2$ chiral pseudofermion actions (\ref{eq:new_MP_N}) (or its parity partner),
while the old MP involves $ 2N+2 $ chiral pseudofermion actions 
(\ref{eq:old_MP_N}) (or its parity partners). 
This implies that the speed-up of the new MP (versus the old MP)   
becomes higher as $ N $ is larger, with the upper bound $ \sim (2N+2)/(N+2) $ for 
a single quark flavor.
This feature may be crucial for lattice QCD simulation in the physical limit 
with a very large volume, in which a cacade of heavy mass preconditoners are required 
to speed up the simulation.

\begin{acknowledgments}
  This work is supported by the Ministry of Science and Technology  
  (Nos.~105-2112-M-002-016, 102-2112-M-002-019-MY3), Center for Quantum Science and Engineering 
  (Nos.~NTU-ERP-103R891404, NTU-ERP-104R891404, NTU-ERP-105R891404),  
  and National Center for High-Performance Computing (NCHC-j11twc00).
  TWC also thanks National Center for Theoretical Sciences for kind hospitality in the summer 2017. 
\end{acknowledgments}


\begin{thebibliography}{99}

\bibitem{Duane:1987de} 
  S.~Duane, A.~D.~Kennedy, B.~J.~Pendleton and D.~Roweth,
  Phys.\ Lett.\ B {\bf 195}, 216 (1987).
  doi:10.1016/0370-2693(87)91197-X

\bibitem{Omelyan:2001}
  I.P.~Omelyan, I.M.~Mryglod, and R.~Folk, Phys.\ Rev.\ Lett. {\bf 86}, 898 (2001).

\bibitem{Sexton:1992nu} 
  J.~C.~Sexton and D.~H.~Weingarten,
  Nucl.\ Phys.\ B {\bf 380}, 665 (1992).
  doi:10.1016/0550-3213(92)90263-B


\bibitem{Hasenbusch:2001ne} 
  M.~Hasenbusch,
  Phys.\ Lett.\ B {\bf 519}, 177 (2001)
  doi:10.1016/S0370-2693(01)01102-9
  [hep-lat/0107019].

\bibitem{Chen:2014hyy} 
  Y.~C.~Chen, T.~W.~Chiu [TWQCD Collaboration],
  Phys.\ Lett.\ B {\bf 738}, 55 (2014)
  doi:10.1016/j.physletb.2014.09.016
  [arXiv:1403.1683 [hep-lat]].

\bibitem{Chen:2014bbc} 
  Y.~C.~Chen, T.~W.~Chiu [TWQCD Collaboration],
  PoS IWCSE {\bf 2013}, 059 (2014)
  [arXiv:1412.0819 [hep-lat]].

\bibitem{Chiu:2015sea} 
  T.~W.~Chiu,
  Phys.\ Lett.\ B {\bf 744}, 95 (2015)
  doi:10.1016/j.physletb.2015.03.036
  [arXiv:1503.01750 [hep-lat]].

\bibitem{Chen:2017kxr} 
  Y.~C.~Chen, T.~W.~Chiu [TWQCD Collaboration],
  Phys.\ Lett.\ B {\bf 767}, 193 (2017)
  doi:10.1016/j.physletb.2017.01.068
  [arXiv:1701.02581 [hep-lat]].

\bibitem{Chiu:2002ir} 
  T.~W.~Chiu,
  Phys.\ Rev.\ Lett.\  {\bf 90}, 071601 (2003)
  doi:10.1103/PhysRevLett.90.071601
  [hep-lat/0209153].

\bibitem{Chiu:2013aaa} 
  T.~W.~Chiu [TWQCD Collaboration],
  J.\ Phys.\ Conf.\ Ser.\  {\bf 454}, 012044 (2013)
  doi:10.1088/1742-6596/454/1/012044
  [arXiv:1302.6918 [hep-lat]].

\end{thebibliography}
\end{document}